\documentclass[conference]{IEEEtran}
\IEEEoverridecommandlockouts
\usepackage{cite}
\usepackage{amsmath,amssymb,amsfonts}
\usepackage{algorithmic}
\usepackage{graphicx}
\usepackage{textcomp}
\usepackage{xcolor}
\usepackage[hidelinks]{hyperref}
\def\BibTeX{{\rm B\kern-.05em{\sc i\kern-.025em b}\kern-.08em
    T\kern-.1667em\lower.7ex\hbox{E}\kern-.125emX}}
\begin{document}

\title{Requirements Debt in AI-Enabled Perception Systems Development: An Industrial RE4AI Perspective\\
\thanks{}
}

\author{\IEEEauthorblockN{1\textsuperscript{st} Hina Saeeda}
\IEEEauthorblockA{\textit{Dept. of IDSE, Computer Science and Engineering  } \\
\textit{Chalmers University of Technology \& University of Gothenburg	}\\
Gothenburg, Sweden \\
hinasa@chalmers.se}
\and
\IEEEauthorblockN{2\textsuperscript{nd} Soniya	Abraham}
\IEEEauthorblockA{\textit{Dept.of Computer Science and Engineering } \\
\textit{Chalmers University of Technology }\\
Gothenburg, Sweden \\
soniyaa@student.chalmers.se}

}

\maketitle

\begin{abstract}
AI integration in automotive perception systems shifts requirements from static specifications to continuously evolving entities shaped by data, models, and operating contexts. When such changes are not consistently documented, validated, and traced, they accumulate as Requirements Debt (ReD), an underexplored but critical subtype of technical debt. This study conceptualises and empirically investigates how evolving functional and non-functional requirements create and propagate ReD across the AI-enabled automotive perception system lifecycle. We conducted 16 semi-structured interviews with experts from 13 international automotive companies and 3 European research institutes and analysed the data using thematic analysis. As one of the first empirical studies connecting technical debt theory with RE4AI, the work identifies key ReD mechanisms. Evolving functional requirements (e.g., algorithm updates, sensor fusion, architectural changes, real-time constraints) drive semantic drift, validation backlogs, and integration debt when verification lags behind rapid iteration. In parallel, evolving non-functional requirements (e.g., safety, cybersecurity, reliability, scalability, transparency, trustworthiness) create assurance lag, compliance misalignment, and transparency and reliability debt as standards and ethical expectations shift. These interacting mechanisms propagate ReD across data, models, and system artefacts, undermining auditability, reliability, and certification readiness in safety-critical perception systems.
\end{abstract}


\section{Introduction}

\textbf{Context.}
Technical Debt (TD) captures the long-term consequences of short-term software development shortcuts~\cite{Kruchten,Tom}. While TD has been widely studied at the architecture, design, and code levels, \textit{Requirements Debt} (ReD) the accumulation of incomplete, ambiguous, inconsistent, or outdated requirements that hinders system evolution~\cite{Melo,Barbosa} remains comparatively underexplored. Existing work offers early classifications and notes the lack of systematic detection and management mechanisms~\cite{Lenarduzzi,wang}, but largely assumes relatively stable requirements typical of traditional software development.

\textbf{Motivation.}
AI integration in safety-critical cyber-physical domains such as automotive perception fundamentally changes requirements from static specifications to evolving constructs shaped by data, model behavior, and operational context~\cite{Saeeda,Peng,Ahmad,Heyn,Habibullah2023REFSQ}. Unlike deterministic systems, AI-enabled Automotive Perception Systems depend on probabilistic models and must satisfy stringent non-functional requirements (e.g., safety, fairness, transparency, privacy) under uncertainty~\cite{Mehrabi2021,NFR4ML2023}. Development proceeds through continuous cycles of data collection, annotation, retraining, and validation, increasingly constrained by ethical and regulatory standards (e.g., ISO 26262 and the EU AI Act)~\cite{Heyn,Habibullah2023REFSQ,ISO26262,EUAIAct}. Consequently, evolving requirements are inherent to AI system engineering, but they also create conditions outdated requirements, inconsistencies, and traceability gaps that are symptomatic of ReD~\cite{Melo,Lenarduzzi}.

\textbf{Research Gap.}
Despite growing attention to RE4AI, limited empirical and conceptual understanding exists on how evolving requirements drive the accumulation and propagation of ReD across the AI-enabled automotive perception systems development lifecycle~\cite{Vogelsang2019,StatusQuo2023}. Changes in datasets, objectives, or performance thresholds can invalidate prior requirements or introduce new ones~\cite{Heyn,Saeeda}; when poorly synchronized across lifecycle stages (from data acquisition and labelling to deployment and monitoring), these changes can cascade and undermine reliability, safety, and trustworthiness~\cite{Habibullah2023REFSQ,Heyn,ISO26262}. The lack of grounded explanations of these causal mechanisms limits organisations’ ability to manage requirement-level risks in AI-enabled systems~\cite{Hatcliff2014}.

\textbf{Research Aims.}
Building on evidence that evolving requirements are a major cause of ReD~\cite{wang}, this study conceptualizes how evolving requirements contribute to the formation and accumulation of ReD in AI-enabled automotive perception systems development.

\textbf{Research Questions.}
\begin{itemize}
\item \textbf{RQ1:} How do evolving or changing requirements contribute to the accumulation and propagation of requirements debt in AI-enabled automotive perception systems?
\item \textbf{RQ2:} How do evolving or changing non-functional requirements contribute to the accumulation and propagation of requirements debt in AI-enabled automotive perception systems?
\end{itemize}

\textbf{Methodology.}
We conducted 16 semi-structured interviews with 13 practitioners and 3 researchers from 13 international automotive companies and 3 research institutes involved in AI-enabled perception system development. The study applies multi-level triangulation by integrating qualitative case study evidence, thematic analysis, and literature review~\cite{Runeson2009,Yin2018}.

\textbf{Novelty and Contribution.}
This work is among the first empirical studies to connect technical debt theory with RE4AI in the automotive context. It (i) identifies evolving requirements as a key source of ReD in AI-enabled automotive perception systems, and (ii) explicates the causal mechanisms through which evolving requirements accumulate and propagate ReD across the AI lifecycle.

This study advances RE4AI toward \textit{debt-aware}, \textit{trustworthy}, and \textit{sustainable} engineering of AI-enabled automotive perception systems~\cite{ISO26262,EUAIAct,Hatcliff2014}.

\section{Background and Related Work}

\textbf{Automotive Perception Systems.}
Automotive perception systems enable vehicles to sense and interpret their surroundings by fusing camera, LiDAR, radar, ultrasonic, and GPS data, forming the basis of ADAS/ADS~\cite{Szeliski2022,Koopman2017,Memon2022}. As data- and model-driven components, they must continuously adapt to new data and conditions, which complicates RE and can accumulate \textit{requirements debt} (ReD) when assumptions and targets evolve~\cite{Heyn,Habibullah2023REFSQ,Heyn2021WAIN,Vogelsang2019}.

\textbf{AI in Automotive Perception Systems.}
AI advances have strengthened perception capabilities such as detection, tracking, and distance estimation, often through multi-sensor fusion~\cite{Heyn2021WAIN,Habibullah2023REFSQ,Skruch2022,Tang2023Fusion}. These \textit{AI-enabled automotive perception systems} are safety-critical and therefore demand high accuracy, reliability, and explainability~\cite{ISO26262,Ballingall2023,Kochanthara2024}.

\textbf{Requirements Engineering for Safety-Critical AI Systems.}
Safety-related development is guided by standards such as \emph{ISO 26262}~\cite{Martins2016SLR,ISO26262}, yet specifying safety requirements remains challenging due to interdisciplinarity, ambiguity risks, and fast-changing technologies and regulations~\cite{book1,Liebel2018,Bjarnason2011,Hatcliff2014}. AI further introduces difficult-to-verify NFRs (e.g., fairness, transparency, privacy, security) in probabilistic, context-dependent models~\cite{NFR4ML2023,Mehrabi2021,Habibullah2023REFSQ,Kochanthara2024}.

\textbf{Evolving Requirements in AI-Enabled Systems Development.}
AI development inherently involves iterative cycles of data collection, annotation, retraining, and validation, making requirements continuously evolving~\cite{Heyn,Habibullah2023REFSQ,wang}. In automotive perception, co-evolution of data, models, sensors, and governance (e.g., EU AI Act, ISO 26262) amplifies this effect~\cite{EUAIAct,ISO26262,StatusQuo2023}. Poor synchronization of requirement changes across lifecycle stages creates inconsistencies and traceability gaps that drive ReD~\cite{Melo,Lenarduzzi,wang}.

\textbf{Requirements Debt (ReD).}
ReD is a form of technical debt caused by incomplete, ambiguous, inconsistent, or outdated requirements, often due to time pressure, weak communication, and shifting goals~\cite{Melo,Lenarduzzi,Bjarnason2011,Liebel2018}. Deferred clarification, validation, or traceability leads to long-term rework and quality degradation~\cite{Tom,Kruchten}.

\textbf{ReD in AI-Enabled Systems.}
In AI contexts, ReD emerges and propagates as data distributions shift, model behavior changes, and new regulatory/ethical expectations introduce additional NFRs~\cite{wang,StatusQuo2023,ISO26262,EUAIAct,Mehrabi2021}. When such changes are not consistently captured from data collection to deployment, ReD accumulates and undermines reliability and trustworthiness~\cite{Habibullah2023REFSQ,Peng,Liu2022}.

\textbf{ReD Implications for RE4AI.}
Because requirements, data, and models are tightly coupled, ReD becomes a lifecycle-wide challenge requiring continuous monitoring and coordination core concerns of \textit{RE4AI}~\cite{Vogelsang2019,Heyn2021WAIN,ISO23053}. Understanding how evolving requirements drive ReD is essential for sustaining trustworthy, compliant, safety-critical automotive perception systems~\cite{Hatcliff2014,ISO26262,Habibullah2023REFSQ}.

\textbf{Problem statement.}
Although AVs rely on real-time perception for safe decisions, deployment remains limited due to sensor uncertainty, environmental complexity, and perception failures~\cite{Heyn,Habibullah2023REFSQ,Koopman2017,Kochanthara2024,Skruch2022}. This requires requirements to evolve with real-world conditions, technological progress, and safety/fairness expectations, yet empirical evidence on how such evolution contributes to ReD in perception system development remains scarce~\cite{Heyn2021WAIN,Heyn2022ODD,Mehrabi2021,Habibullah2023REFSQ,Heyn,Ribeiro2022,Vogelsang2019}.

\section{Research Methodology}
\label{sec:methods}

\begin{figure*}[!ht]
    \centering
    \includegraphics[width=\textwidth]{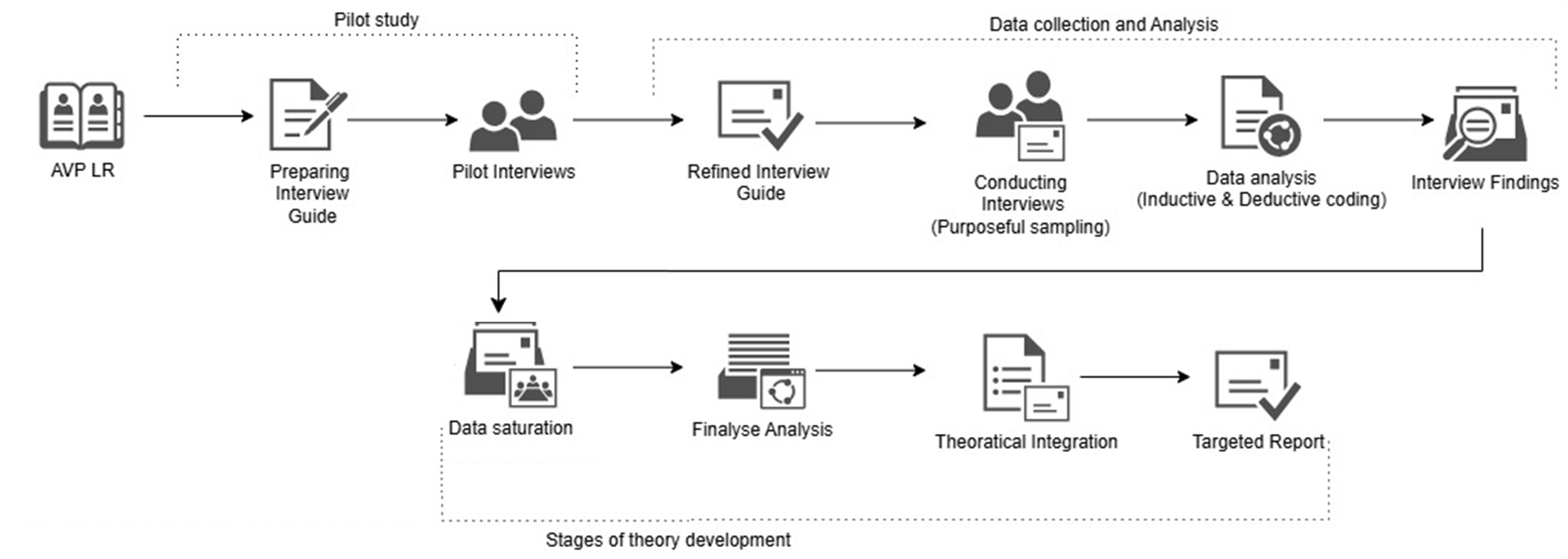}
    \caption{Overview of the research process.}
    \label{fig:Diagram-method}
\end{figure*}

We followed the qualitative case study guidelines by Runeson and Höst~\cite{Runeson2009}, complemented by Yin’s systematic procedures for case study research~\cite{Yin2018}, to ensure methodological rigour. Fig~\ref{fig:Diagram-method} provides an overview of the methodological flow.

\subsection{Sampling and Participants}
We conducted sixteen semi-structured interviews following qualitative interviewing best practices~\cite{McNamara2009}.  
Participants were selected using purposeful sampling~\cite{Stol2018}, ensuring diversity across roles (e.g., architects, engineers, researchers) and organizational tiers involved in AI-enabled Perception Systems (AIePS) development. This included representatives from OEMs, Tier-1 and Tier-2 suppliers, and research institutes. Interviews were performed online between March and June 2025, each lasting approximately 60 minutes.  
All interviews were recorded, transcribed, and validated through participant feedback, consistent with ethical research standards~\cite{Runeson2009,Yin2018}.  

\begin{table}[!ht]
\centering
\scriptsize
\setlength{\tabcolsep}{3pt}
\renewcommand{\arraystretch}{1.1}
\caption{Interviewee details}
\label{tab:interviewees}
\begin{tabular}{|c|l|l|c|}
\hline
\textbf{ID} & \textbf{Role} & \textbf{Company} & \textbf{Exp. (yrs)} \\ \hline
ID1  & CEO                & Zion Tech AB            & 17+ \\ \hline
ID2  & Software Engineer   & Univrses                & 4   \\ \hline
ID3  & Senior Engineer     & Polestar                & 13  \\ \hline
ID4  & Engineer            & Kognic                  & 5   \\ \hline
ID5  & Solution Architect  & Volvo                   & 13  \\ \hline
ID6  & Researcher          & Gothenburg University   & 2+  \\ \hline
ID7  & Architect           & Zenseact                & 4   \\ \hline
ID8  & ML Engineer         & Cruise                  & 5+  \\ \hline
ID9  & Engineer            & Wabtec                  & 6   \\ \hline
ID10 & Researcher          & RISE                    & 2+  \\ \hline
ID11 & Associate Professor & Halmstad University     & 7   \\ \hline
ID12 & Engineering Manager & Aptiv                   & 17  \\ \hline
ID13 & System Designer     & Zenseact                & 5+  \\ \hline
ID14 & Engineer            & Capgemini               & 3   \\ \hline
ID15 & Test Architect      & Volvo                   & 5   \\ \hline
ID16 & Architect           & Scania                  & 13  \\ \hline
\end{tabular}
\end{table}
\subsection{Organizational Context}
Our participants represented thirteen organisations spanning the AI-enabled automotive perception systems development ecosystem, covering one OEM, several Tier-1 suppliers (requirements and safety-critical software), and Tier-2 companies (annotation, perception, and data services).  
Three experts from academia and research institutes were also included to provide cross-sectoral validation.  
This diversity ensured a holistic view of evolving requirements, challenges, and cause-effect analysis strategies across the full AI system development lifecycle practices in the Swedish Automotive industry~\cite{Heyn,Habibullah2023REFSQ}.

\subsection{Interview Design and Preparation}
A semi-structured interview guide was iteratively developed based on prior RE4AI studies~\cite{Heyn2021WAIN,Vogelsang2019,Habibullah2023REFSQ}.  
Two researchers co-designed and pilot-tested the guide for clarity and content validity.  Two pilot interviews were conducted with domain-familiar practitioners selected from the researchers' professional networks. These were excluded from the final analysis.
The guide included open-ended questions targeting (i) evolving requirements, (ii) challenges, (iii) causes, and (iv) consequences ~\cite{Steghofer2019,Dikert2016}.  
Questions were mapped to the study’s research questions and informed by the theoretical framing from prior work~\cite{Heyn,Liebel2018}. For transparency, the final interview guide is publicly available (see \href{https://dataverse.harvard.edu/file.xhtml?fileId=13091726&version=1.0}{\textcolor{blue}{→ Interview Guide}}).

\subsection{Data Analysis}
We employed Braun and Clarke’s six-phase Thematic Analysis framework~\cite{them}, complemented by Saldaña’s coding techniques for qualitative research~\cite{coding}.  
This process included (1) familiarization with data, (2) generating initial codes, (3) searching for themes, (4) reviewing themes, (5) defining and naming themes, and (6) reporting.  
A hybrid deductive inductive approach was used, where initial codes were informed by the research questions and a code book was designed based on these questions (see:\href{https://dataverse.harvard.edu/file.xhtml?fileId=13091725&version=1.0}{\textcolor{blue}{→ Codebook}}). Although interview questions were grounded in identifying critical evolving requirements, participants consistently described these requirements as shifting over time due to AI model updates, sensor changes, and regulatory evolution. This emergent emphasis on requirement volatility, combined with descriptions of deferred validation, documentation gaps, and traceability failures, provided the empirical basis for inductively coding ReD mechanisms. The ReD constructs reported in Section IV are thus grounded in participant descriptions of specific failure modes, not only imposed from prior literature, though they were subsequently connected to the ReD conceptual frameworks from literature for theoretical integration. The codes were expanded inductively as new patterns emerged.
Manual coding was performed in Excel (see:\href{https://dataverse.harvard.edu/file.xhtml?fileId=13091730&version=1.0}{\textcolor{blue}{→ Transcripts Analysis and Codes Extraction}}).
Visual clustering was done on MIRO boards for traceability and transparency.
Researcher triangulation was applied through independent coding and calibration meetings~\cite{Runeson2009}, enhancing interpretive reliability.  
In total, sixteen transcripts were analysed, yielding over 800 coded text segments and five overarching themes with eighteen sub-themes describing evolving requirements, challenges, and consequences in AIePS development.
(see:\href{https://dataverse.harvard.edu/file.xhtml?fileId=13091729&version=1.0}{\textcolor{blue}{→ Identified Codes and Sub-Themes }}).

\subsection{Ethical Considerations}
All participants provided informed consent. Interviews were anonymized, securely stored, and handled in accordance with ethical research standards for empirical software engineering~\cite{Runeson2009,Yin2018}.  
Sensitive company and participant identifiers were removed to maintain confidentiality, and data handling complied with GDPR and institutional ethical guidelines.

\section{Findings }

\subsection{How do evolving or changing functional requirements contribute to the accumulation and propagation of requirements debt in AI-enabled automotive perception systems development ?}

\begin{figure*}[!ht]
    \centering
    \includegraphics[width=0.80\textwidth]{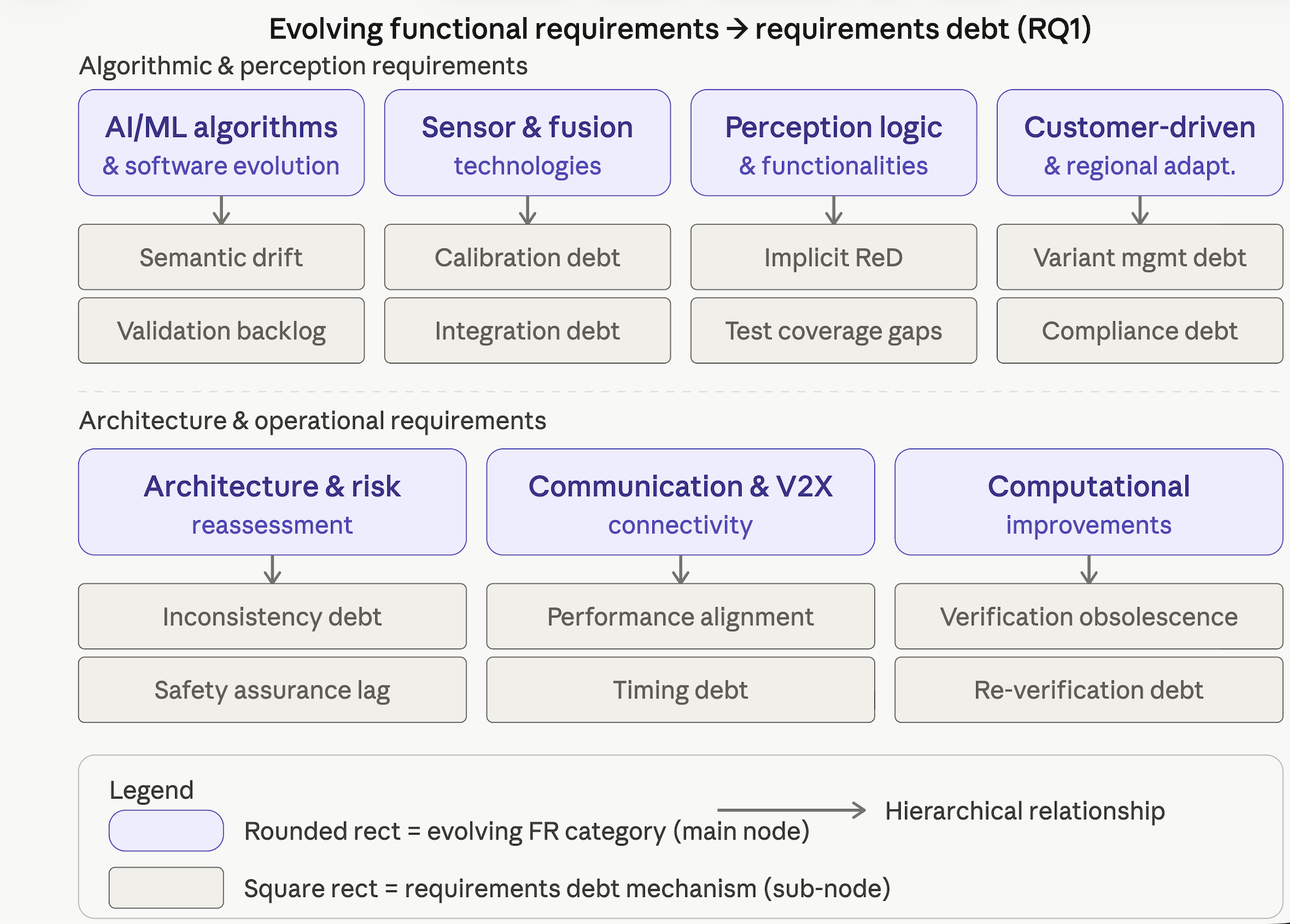}
    \caption{Identified Evolving Functional Requirements Leading to ReD}
    \label{fig:functional}
\end{figure*}

Evolving functional requirements in AI-enabled perception systems (Fig.~\ref{fig:functional}) driven by rapid advances in AI/ML, sensor technologies, and connectivity are a major source of requirements debt (ReD). While these changes improve capability and safety, they often outpace documentation, traceability, and validation, creating inconsistencies and backlogs that accumulate and propagate ReD. The following synthesis summarises how the interview-identified requirement categories contribute to these debt mechanisms.

\subsubsection{Evolving Technology and Algorithmic Complexity}

As participants emphasised, autonomous systems have shifted from rule-based logic toward AI-driven perception and prediction, relying on data-hungry and continuously evolving models. One interviewee stated:
\textit{“To be able to have a good perception system, you need to be able to track objects in a 3D world… and also make predictions... where is this object gonna be in a certain period of time?”} (ID10).

Such dynamic learning capabilities require ongoing redefinition of functional requirements, accuracy thresholds, prediction horizons, and acceptable uncertainty margins, which often lag behind implementation. When requirements specifications fail to keep pace with new AI capabilities, semantic drift debt emerges: the documented requirements no longer match what the system actually performs. Moreover, as another participant noted,
\textit{“Software is becoming the key differentiator... even the same hardware can behave differently because of new software.”} (ID6).
This observation highlights how frequent software updates driven by model retraining or algorithm enhancement lead to validation debt deferred or incomplete verification of new behaviours introduced through AI software evolution.

\textbf{Challenge caused by evolving requirements:} Rapid AI/ML evolution creates \emph{semantic drift} between documented requirements and actual learned behaviour (new prediction horizons, uncertainty budgets, decision logic).
\textbf{Leads to ReD via:} Outdated specifications and acceptance criteria, plus a validation backlog as software updates outpace re-certification.

\subsubsection{Advancing Sensor Technologies and Fusion Requirements}

Participants repeatedly described how rapid improvements in cameras, LiDAR, and radar reshape system design:
\textit{“We’re moving beyond drive-by-wire systems, evolving hardware with new adaptive sensors like cameras, LiDAR, and thermal imaging.”} (ID1).
Each hardware upgrade changes core parameters, resolution, range, or field of view that underpin the perception system’s requirements. When these updated specifications are not fully propagated through architectural design or testing documents, interface and calibration debt accumulates. For instance, perception fusion models must be recalibrated whenever a sensor specification changes; skipping or postponing this step introduces inconsistencies between data inputs and the fusion logic, degrading overall reliability.

Furthermore, as perception pipelines integrate heterogeneous sensors through deep learning, requirement dependencies become nonlinear one small parameter change in LiDAR precision can ripple across multiple requirement layers. This cross-dependency amplifies integration debt, in which unclear traceability between sensor-level requirements and AI model behaviour increases maintenance costs and re-validation effort.

\textbf{Challenge caused by evolving requirements:} Fast sensor upgrades drive \emph{interface \& calibration churn} and nonlinear \emph{fusion dependencies}.
\textbf{Leads to ReD via:} Partial propagation of new sensor requirements, fusion inconsistencies, and growing integration/traceability gaps across versions.

\subsubsection{Evolving Functionalities and Perception Logic}

Participants emphasised that perception systems evolve from detecting static objects to predicting dynamic trajectories in uncertain environments:
\textit{“Perception systems require a combination of object detection, recognition, tracking, and prediction, with AI playing a key role.”} (ID10).
Each new capability (e.g., trajectory forecasting, intent prediction) introduces a set of implicit functional requirements: temporal accuracy, prediction confidence, and reaction latency that may not be explicitly updated in requirement baselines. These latent expectations form implicit requirements debt, where undocumented but assumed functionalities later lead to inconsistencies in validation and testing. As AI-enabled systems' perception becomes predictive and context-aware, the mismatch between evolving functionality and static requirements documentation compounds long-term ReD.

\textbf{Challenge caused by evolving requirements:} New AI-driven capabilities introduce \emph{implicit/latent requirements} that are not formalised in time.
\textbf{Leads to ReD via:} Missing acceptance criteria and test coverage, causing downstream inconsistencies and rework in validation.

\subsubsection{Customer-Driven and Region-Specific Adaptations}

Manufacturers face the continuous need to adapt to customer expectations and regional standards. One participant observed,
\textit{“User expectation is broadly fixed… but new expectations like watching Netflix or playing games can emerge as customer demands evolve.”} (ID9).
These market-driven shifts necessitate ongoing modifications to requirements that often lack systematic version control, resulting in variant management debt. Similarly, evolving legal frameworks such as Euro NCAP and country-level homologation standards demand regional customisation. As another expert noted,
\textit{“At the beginning, it was only Euro NCAP… nowadays, we have started to see that these tests are being fine-tuned for each and every country.”} (ID12).
Such fragmentation generates compliance debt, as different versions of requirements must coexist across markets without a unified traceability structure. This complicates impact analysis when perception models or architectures change globally.

\textbf{Challenge caused by evolving requirements:} Market features and regional rules create \emph{variant fragmentation} and shifting compliance targets.
\textbf{Leads to ReD via:} Divergent baselines without unified traceability, duplicated testing, and complex impact analysis across variants.

\subsubsection{Architectural Evolution and Risk Reassessment}

The transition from static architectures to adaptive, end-to-end transformer-based frameworks was repeatedly highlighted:
\textit{“We need to break away from the drive-by-wire kind of architectures and move more into evolving architectures where you take decisions based on situations.”} (ID1).
These architectural shifts require corresponding updates in performance, safety, and fault-tolerance requirements. However, architectural evolution often precedes revisions to formal requirements, resulting in architectural debt, where legacy requirements constrain or contradict the capabilities of modern AI-based architectures.

In parallel, risk assessment frameworks lag behind evolving methodologies. As stated by one participant,
\textit{“As the tools are evolving and the methodology is changing, the risk analysis will give different scores... we have to adapt our testing and development strategies.”} (ID2).
When risk and testing strategies evolve asynchronously with system updates, safety assurance debt accumulates, jeopardizing validation completeness and system certification readiness.

\textbf{Challenge caused by evolving requirements:} \emph{Architectural misalignment} (legacy requirements vs.\ adaptive/e2e designs) and \emph{safety assurance lag} (risk models trailing new methods).
\textbf{Leads to ReD via:} Contradictory requirements, outdated hazards/evidence, and certification delays.

\subsubsection{Communication, Connectivity, and Real-Time Processing}

Connectivity (V2X) and real-time data processing were frequently identified as emerging requirements.
\textit{“Communication and connectivity are evolving with the increasing need for data exchange between vehicles and infrastructure.”} (ID6).
Another participant noted:
\textit{“V2X domain… is still transitioning and not yet fully implemented in all vehicles… Gen AI is being integrated into most technologies.”} (ID14).
These continuous connectivity enhancements require updated latency, bandwidth, and security requirements. When legacy requirements fail to account for new V2X architectures or Gen-AI-assisted decision frameworks, performance alignment debt arises. Similarly, as one participant stressed,
\textit{“The reaction time is calculated based on different scenarios, including speeds and how long it takes for the vehicle to stop, ensuring that these decisions are made in real-time.”} (ID6).
Delays in updating latency and real-time response requirements following computational improvements lead to timing debt, undermining safety-critical performance guarantees.

\textbf{Challenge caused by evolving requirements:} Shifting V2X/Gen-AI capabilities and strict latency bounds cause \emph{performance misalignment} and \emph{timing drift}.
\textbf{Leads to ReD via:} Miscalibrated latency/throughput/security requirements and erosion of real-time guarantees.

\subsubsection{Computational Improvements and Re-Verification Delays}

Emerging paradigms such as quantum and high-performance edge computing amplify both capability and complexity. As an interviewee remarked,
\textit{“You move from binaries into quantum in the future.”} (ID1).
While such advancements enhance model capability, they simultaneously render prior verification results obsolete, demanding new validation frameworks for correctness, determinism, and explainability. When these re-verifications are deferred, verification debt accumulates, posing long-term risks for certification and trust in AI-enabled perception systems per performance.

\textbf{Challenge caused by evolving requirements:} New compute paradigms create \emph{verification obsolescence} (old evidence no longer valid).
\textbf{Leads to ReD via:} Accumulated re-verification needs, slowed releases, and uncertainty in correctness/explainability.

\subsection{RQ2.How do evolving or changing non-functional requirements contribute to the accumulation and propagation of requirements debt in AI-enabled automotive perception systems?}

\begin{figure*}[!ht]
    \centering
    \includegraphics[width=0.70\textwidth]{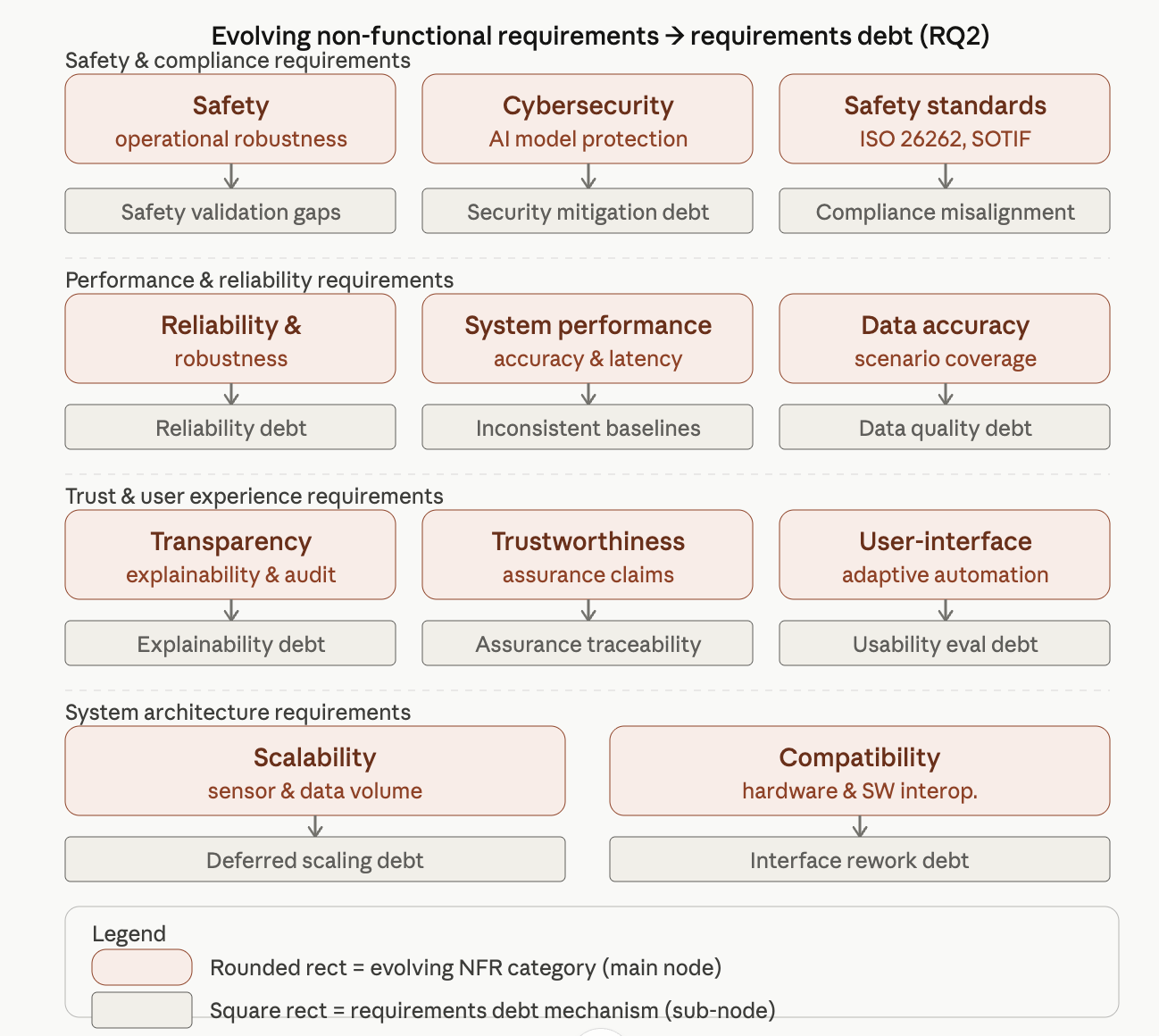}
    \caption{Identified Evolving Non-Functional Requirements Leading to ReD}
    \label{fig:nonfunctional}
\end{figure*}

 This section presents the evolving non-functional requirements (illustrated in Fig~\ref{fig:nonfunctional}) for AI-enabled perception systems, emphasising aspects such as safety, system performance,  user trust, cybersecurity, reliability, scalability, and transparency, which are critical to ensuring that AV systems operate not only effectively but also securely. 

\subsubsection{Safety}

In the context of  AI-enabled perception systems for autonomous vehicles, safety is universally regarded as one of the most critical aspects. Various participants emphasise that safety is a broad and evolving requirement, with different facets contributing to the overall system's ability to operate securely and predictably in real-world conditions. The participant points out the continuous evolution of safety standards, which require vehicles to adapt to new data and scenarios to enhance their capabilities and prevent failures. This constant data collection and validation ensure that the vehicle’s safety mechanisms are continuously tested. One participant states that \textit{“Safety is definitely a crucial requirement. It involves that the vehicle be robust, work as intended, and be able to be resilient.”}(ID6)

\textbf{Challenge caused by evolving requirements:} As safety standards evolve and the system must continuously adapt to new data, ensuring that safety validations remain aligned with emerging risks and scenarios becomes difficult.  
\textbf{Leads to ReD via:} Accumulated gaps in safety validation and documentation, as legacy safety checks may not fully represent current operational conditions.

\subsubsection{Cybersecurity}

Cybersecurity is increasingly crucial in AI-enabled perception systems for autonomous vehicles, as highlighted by some participants, who stress its importance in protecting against potential hacks that could alter system behaviour. The growing attention to cybersecurity is necessary to ensure reliability and trust in autonomous vehicles. In essence, cybersecurity is a critical quality requirement, but its implementation may overlap with functional needs depending on the specific security measures. Multiple participants raised their opinion on cybersecurity, as one of the participants says it, \textit{ "Cybersecurity is getting quite more attention...deep learning models are known to, you know, could be manipulated in certain ways if certain weights are modified..."} (ID2)

\textbf{Challenge caused by evolving requirements:} Increasing AI-enabled perception system connectivity and AI model exposure raise cybersecurity risks that evolve faster than protective mechanisms.  
\textbf{Leads to ReD via:} Unaddressed or outdated security requirements that fail to mitigate new attack surfaces, resulting in deferred mitigation and verification debt.

\subsubsection{Safety standards}

The safety of autonomous vehicles is closely tied to adherence to various standards and regulations. In addition to the SOTIF (ISO 21448:2022) requirements, which focus on ensuring the robustness of systems under adverse conditions such as bad weather or hazardous scenarios, other key standards include ISO 26262 for functional safety. This standard is fundamental in ensuring that the vehicle operates safely in all situations, especially when unforeseen events or malfunctions occur. Furthermore, ISO 21448 addresses the system's functional safety in non-critical situations. As one participant highlighted, \textit{ ``SOTIF actually stands for Safety of the Intended Functionality. So it describes how the intended function can behave in a hazardous way when there are adverse situations."} (ID16)

\textbf{Challenge caused by evolving requirements:} Rapid evolution of standards (e.g., ISO 26262, ISO 21448, SOTIF) creates continual updates to compliance targets that may not be reflected across all design and validation documents.  
\textbf{Leads to ReD via:} Compliance misalignment and outdated certification evidence that require retrospective re-validation.

\subsubsection{Reliability and robustness}

Reliability and robustness are key pillars in the evolving landscape of autonomous vehicles, especially as customer expectations grow and vehicles operate in increasingly unpredictable environments. Robustness is not just about physical durability but also about ensuring consistent performance and fault tolerance when systems encounter edge (e.g., unforeseen situations, undefined scenarios) or environmental challenges. At the same time, reliability ensures that the vehicle consistently performs its functions without failure, earning customer trust over time. One participant says about it \textit{"If you want me to rank or give one name, then I will say robustness, reliability, those are super important, even if it is working at 1 Hz, very slow speed."}-(ID11). 

\textbf{Challenge caused by evolving requirements:} As AVs face new environments and data distributions, maintaining robustness and reliability across scenarios becomes increasingly complex.  
\textbf{Leads to ReD via:} Deferred stress testing and fault-tolerance updates, creating reliability debt and performance gaps in unseen situations.

\subsubsection{System performance}

Some participants place strong emphasis on performance from a machine learning standpoint, highlighting the importance of accuracy metrics across diverse environmental conditions (e.g., day, night, fog) and different types of road users, as well as the role of latency in determining model responsiveness. As the participant puts it, \textit{“The main requirement is to ensure that even if there is a regression, it is justified by an improvement somewhere else that may be more important,” }-ID8

\textbf{Challenge caused by evolving requirements:} Shifts in model performance goals (accuracy vs.\ latency trade-offs) make it difficult to maintain stable baselines across software releases.  
\textbf{Leads to ReD via:} Inconsistent performance thresholds and unclear regression handling policies, requiring later performance recalibration.

\subsubsection{Data accuracy}

Participants discussed the critical role of accuracy in autonomous vehicle performance, particularly in tracking, threat assessment, and decision-making. Accuracy ensures that the vehicle can correctly identify and assess objects in its environment, especially when distinguishing potential threats. It is essential for determining whether a collision is imminent and ensuring appropriate actions are taken, such as braking. As one participant noted, \textit{"The accuracy is calculated with all different scenarios, and then you take the worst scenario. You estimate the accuracy." }(ID5). 

\textbf{Challenge caused by evolving requirements:} As AI-enabled perception systems encounter diverse and unstructured data, maintaining accuracy across scenarios becomes harder.  
\textbf{Leads to ReD via:} Accumulated data quality and calibration debt due to outdated accuracy thresholds and limited coverage of rare cases.

\subsubsection{User-interface}

Participants highlighted that the key aspect of the user experience is ensuring that interactions with AI-enabled perception systems remain intuitive and less distracting for users. As autonomous vehicles advance, users will increasingly expect to focus less on driving and more on personal activities, such as using their phones or preparing for meetings while travelling. The UI is designed to reduce distraction while still providing essential information and controls, creating a seamless balance between automation and user engagement. As one participant noted, \textit{ ``People like to get more comfortable and they don't want to have less focus on driving, but more focus when they travel" }(ID5).

\textbf{Challenge caused by evolving requirements:} As automation increases, evolving user expectations demand adaptive interfaces that balance control, comfort, and awareness.  
\textbf{Leads to ReD via:} Inconsistent interface requirements, outdated usability evaluations, and deferred accessibility validation.

\subsubsection{Transparency}

Interviewees emphasised the importance of transparency in an AI-enabled perception system for autonomous vehicles, particularly regarding data transfer and system predictions. Ensuring that the system's predictions are not only accurate but also explainable and understandable to end users is crucial for fostering trust and confidence in autonomous technologies. Transparency in data communication, including V2X (vehicle-to-Everything) interactions, is equally important for guaranteeing the quality and security of the transmitted information. As one participant highlighted, \textit{"Ensuring security during data transfer and making predictions that are explainable and understandable for end users is also important"} (ID11), 

\textbf{Challenge caused by evolving requirements:} Growing demand for explainable and auditable AI-enabled models creates pressure to make complex ML decisions transparent to users.  
\textbf{Leads to ReD via:} Lack of explainability frameworks and documentation debt in prediction and decision-making transparency.

\subsubsection{Trustworthiness}

Trustworthiness is a critical factor in the development and acceptance of autonomous vehicles. It is essential for developing public confidence that these AI-enabled perception systems are dependable and secure. Security concerns, particularly with the vulnerability of deep learning models to manipulation, make it even more important to address cybersecurity as a central focus. As one participant noted,\textit{``Cybersecurity is getting quite more attention. And of course, reliability is one of the key aspects... to make them more reliable and in general increase the trust in the people as well"} (ID2).

\textbf{Challenge caused by evolving requirements:} Establishing user trust in continuously learning AI-enabled perception systems that change behaviour over time is inherently difficult.  
\textbf{Leads to ReD via:} Incomplete traceability between assurance claims (safety, reliability) and evolving AI-enabled model behaviour.

\subsubsection{Scalability}

Participants are concerned that as technology advances, the introduction of more sensors and cloud integrations poses challenges for vehicle architectures to satisfy these developments. A major challenge is the capacity to handle and understand the huge amounts of data produced by these sensors. Moreover, the requirement for scalability encompasses the ability to adapt to new traffic conditions and to ensure that the vehicle's systems can handle complex scenarios. One participant underlined this issue, remarking, \textit{ ``The scalability aspect is a significant question. And how do you adjust to traffic conditions? Thus, I see these as areas that act as a kind of roadblock for scalability...}(ID1).

\textbf{Challenge caused by evolving requirements:} Increasing sensor density, data volume, and cloud integration raise scalability challenges for AI-enabled perception systems architectures.  
\textbf{Leads to ReD via:} Deferred architectural scaling, resource bottlenecks, and delayed adaptation to high-load conditions.

\subsubsection{Compatibility}

This involves considering the interoperability of various sensors, software packages, and communication systems such as V2X, which enable vehicles to share information with one another and with infrastructure. It is also essential that these systems can be modular, making it easy to upgrade or adjust without affecting the overall system’s functionality. One participant pointed out, \textit{"Of course, compatibility issues also matter, like whether the things developed nowadays are compatible with existing hardware systems, computational systems on the vehicle, or modularity."} (ID11)

\textbf{Challenge caused by evolving requirements:} Rapid hardware/software evolution leads to \emph{compatibility drift} across platforms and suppliers.  
\textbf{Leads to ReD via:} Integration rework, interface mismatches, and delayed modular upgrades as systems evolve independently.

\section{Threats to Validity}

Following Runeson~\textit{et al.}~\cite{Runeson2009}, we outline validity threats and mitigations.

\textbf{Construct Validity.}
We purposefully sampled relevant experts from OEMs, Tier-1 and Tier-2 suppliers, and research institutes. The interview protocol was internally reviewed and pilot-tested twice. Key concepts (e.g., evolving requirements, perception systems) were defined and supported with visuals and pre-interview briefings to reduce misinterpretation. \textbf{(See Replication Data).}
\footnote{\href{https://doi.org/10.7910/DVN/YBEUKX}{\textcolor{blue}{\textit{Supplementary material available here}}}.}
Researcher bias was mitigated through reflexive team discussions and peer debriefing.

\textbf{Internal Validity.}
To reduce individual and organisational bias, participants were recruited across 13 automotive companies and 3 research institutes. Interviews were recorded, transcribed, and participant-validated. Two researchers coded independently ($\kappa = 0.8$), resolving disagreements in recurring meetings (March--June 2025) while iteratively refining themes.

\textbf{External Validity.}
While not statistically generalisable, the study supports analytical generalisation through diversity across 16 organisations in European contexts. Coverage across supply-chain tiers and alignment with relevant standards (ISO/IEC~5259, IEEE~P2801, ISO~26262, SAE~J3016) strengthen transferability.

\section{Discussion}

\subsection{Study Overview and Positioning within RE4AI and Technical Debt}
This study explains how evolving functional and non-functional requirements in AI-enabled perception systems (AIePS) accumulate as requirements debt (ReD) in the automotive domain. It strengthens RE4AI by extending ReD beyond relatively stable, traditional requirements contexts~\cite{Ahmad,Saeeda,Heyn2021WAIN} and broadens technical debt discourse to data-centric, continuously learning systems~\cite{Kruchten,Tom,Lenarduzzi,Melo,Barbosa}. Our results show that requirement evolution is a structural property of the AI lifecycle; when documentation, validation, and traceability do not co-evolve with data, models, and standards, new debt forms emerge~\cite{Habibullah2023REFSQ,Vogelsang2019,NFR4ML2023}.

\subsection{Parallel and Contrast with State-of-the-Art Literature}
Consistent with prior work on uncertainty in data and annotation ecosystems~\cite{Heyn,Heyn2022ODD,Peng}, our findings show how these uncertainties translate into persistent ReD. We identify \textit{semantic drift} as a concrete mechanism linking documented requirements to shifting learned behaviour~\cite{Ahmad,StatusQuo2023}. In contrast to algorithmic debt studies, we show how incremental functional upgrades (e.g., sensor fusion changes) create unverified dependencies across the data--model--requirement triad~\cite{Tang2023Fusion}. For non-functional requirements, we move beyond viewing standards (e.g., ISO 26262/SOTIF) as fixed targets by evidencing \textit{safety assurance lag} driven by evolving interpretations and operational feedback~\cite{ISO26262,Henriksson2018,Skruch2022,Kochanthara2024}. We further connect transparency and explainability to RE practice by characterising \textit{transparency debt} when explainability requirements become outdated and unverifiable~\cite{Liu2022,Mehrabi2021}.

\subsection{Novelty in Comparison with State-of-the-Art Literature}
The key novelty is the empirical mapping of evolving requirements into explicit ReD pathways in AIePS. Beyond recognising volatility as a challenge~\cite{Stol2018}, we characterise traceable debt types (e.g., semantic drift, safety assurance lag, validation backlog, calibration debt, transparency debt) and position ReD as \textit{multi-artefact debt} spanning requirements, data, and models~\cite{Ahmad,Vogelsang2019}. The study also highlights socio-technical drivers of propagation fragmented ownership, delayed communication, and unclear accountability in cross-disciplinary AI ecosystems~\cite{Liebel2018,Bjarnason2011,Steghofer2019}.

\subsection{Implications for Practice and Future Research}

Industry should treat requirements as evolving assets that remain synchronised with datasets, models, and regulation~\cite{Habibullah2023REFSQ,ISO26262}. Embedding Continuous RE into CI/CD and MLOps can keep requirement baselines and validation evidence current~\cite{Heyn2021WAIN,Vogelsang2019}. In practice, organisations can use \textit{debt registers} to track pending re-validations and compliance gaps, adopt adaptive safety cases aligned with retraining cycles and dataset shift~\cite{Hatcliff2014,Ballingall2023}, and strengthen coordination across RE, data, and safety roles~\cite{Liebel2018,Steghofer2019}. Tooling for explainable, versioned traceability across data--model--requirement artefacts can further reduce verification and transparency debt while improving auditability~\cite{NFR4ML2023,Liu2022}.

For research, future work should model and quantify ReD propagation and its impact on reliability and maintainability~\cite{Vogelsang2019,Ahmad}, including longitudinal estimates of ReD ``interest rates'' and repayment costs~\cite{Kruchten,Tom}. An important direction is to investigate the role of \textit{process debt} as an antecedent to ReD, building on prior work \cite {PD}. Process-level inefficiencies, including inadequate documentation, coordination gaps, and time pressure, may create conditions under which requirements are inconsistently specified, insufficiently validated, or poorly traced. Understanding how such process debt propagates into ReD can provide a more holistic, socio-technical perspective on debt accumulation in AI-enabled systems. Future research should also investigate how organisational and coordination challenges in large-scale agile settings influence the emergence and propagation of ReD \cite{LSAD}.

Promising directions include defining measurable indicators (e.g., model--requirement alignment or data-coverage divergence) for early detection, examining how regulation such as the EU AI Act can incorporate ReD monitoring into assurance and certification~\cite{EUAIAct}, and conducting socio-technical studies on how organisational structures and process maturity influence both process debt and ReD recognition and resolution~\cite{Liebel2018,Bjarnason2011}. Longitudinal and multi-organisational studies are particularly needed to capture the dynamic interplay between process practices, evolving requirements, and system-level outcomes in safety-critical AI contexts.

\subsection{Lifecycle Implications for AI-Enabled System Development}
ReD manifests across the full AI lifecycle~\cite{Ahmad,Heyn,Habibullah2023REFSQ}. In data stages, evolving coverage and labeling requirements create early data-related debt that affects generalisation~\cite{Heyn,Peng}. During model development, rapid retraining and architectural changes can outpace requirement updates, producing validation and calibration debt~\cite{Vogelsang2019,Tang2023Fusion}. In integration/testing, delayed update propagation introduces traceability debt. In operation, real-world feedback exposes new constraints; postponed requirement updates yield safety-assurance lag and performance-misalignment debt~\cite{Skruch2022,Kochanthara2024}. Embedding debt awareness through adaptive documentation, automated traceability, and version-controlled artefacts supports verifiability and audit readiness in continuously evolving AI systems~\cite{ISO23053,Hatcliff2014}.

\subsection{Lessons Learned for Debt-Aware RE4AI}
Several lessons emerged from this study. First, managing requirements in AI-enabled perception systems cannot rely on static RE practices. Rapid evolution in data, models, and standards demands a dynamic, adaptive mindset where requirements are continually validated against real-world behavior~\cite{Ahmad,Vogelsang2019}. Second, ReD does not result solely from negligence or poor documentation but from the speed mismatch between evolving AI artefacts and slower RE processes~\cite{Heyn2021WAIN,Saeeda}. This positions ReD as an intrinsic by-product of innovation rather than a management failure. Recognising ReD early, documenting its presence, and prioritising management through iterative refinement are essential for sustainable AI development. Finally, effective RE4AI requires a cultural shift from compliance-driven requirement management to proactive, debt-aware engineering~\cite{Stol2018,Kruchten}. By embedding continuous validation and explainable traceability, organizations can transform ReD from a hidden liability into a structured mechanism for learning and improvement~\cite{NFR4ML2023,Liu2022}.

\subsection{Quality Degradation Impact of the Identified Requirements Debts}
The accumulation of requirements debt contributes directly to multidimensional quality degradation in AI-enabled systems. At the product level, semantic drift and calibration debt erode accuracy and reliability, increasing false positives and missed detections in safety-critical contexts~\cite{Memon2022,Szeliski2022}. Validation backlog and safety-assurance lag weaken testing completeness and delay certification~\cite{ISO26262,Ballingall2023}. Transparency and traceability debt reduce accountability by obscuring rationale chains and hindering post-incident analysis~\cite{Liu2022}. At the process level, unresolved ReD slows development velocity, raises rework costs, and decreases maintainability~\cite{Tom,Kruchten}. At the organizational level, repeated misalignments between RE, data, and safety teams diminish cross-disciplinary trust, creating communication debt and decision latency~\cite{Liebel2018,Bjarnason2011}. These cascading effects establish a feedback loop of quality erosion: as ReD grows, the cost and complexity of verification rise, amplifying risk. Addressing these effects demands a debt-aware governance model where quality assurance, safety validation, and requirements evolution are treated as interdependent processes. Continuous monitoring of ReD indicators can serve as an early-warning mechanism for performance or safety degradation, preserving trustworthiness throughout the AI lifecycle~\cite{Koopman2017,ISO23053}.

\end{document}